\documentclass{amsart}
\usepackage{graphicx}
\usepackage{amscd}
\usepackage{amsmath}
\usepackage{amsfonts}
\usepackage{amssymb}

\newtheorem{theorem}{Theorem}
\theoremstyle{plain}
\newtheorem{acknowledgement}{Acknowledgement}

\newtheorem{case}{Case}
\newtheorem{claim}{Claim}

\newtheorem{corollary}{Corollary}

\newtheorem{definition}{Definition}

\newtheorem{lemma}{Lemma}

\newtheorem{proposition}{Proposition}

\numberwithin{equation}{section}

\begin{document}
\title{{\large Frequency spanning homoclinic families. }}
\author{Vered Rom-Kedar.}
\address{Department of computer science and applied mathematics, Weizmann Institute.}
\email{vered@wisdom.weizmann.ac.il}
\urladdr{http://www.wisdom.weizmann.ac.il/\symbol{126}vered/}
\date{\today}

\begin{abstract}
A family of maps or flows depending on a parameter $\nu$ which
varies in an interval, spans a certain property if along the
interval this property depends continuously on the parameter and
achieves some asymptotic values along it. We consider families of
periodically forced Hamiltonian systems for which the
appropriately scaled frequency $\overline{\omega}(\nu)$ is
spanned, namely it covers the semi-infinite line $[0,\infty).$
Under some natural assumptions on the family of flows and its
adiabatic limit, we construct a convenient labelling scheme for
the primary homoclinic orbits which may undergo a countable number
of bifurcations along this interval. Using this scheme we prove
that a properly defined flux function is $C^{1}$ in $\nu.$
Combining this proof with previous results of RK and Poje,
immediately establishes that the flux function and the size of the
chaotic zone depend on the frequency in a non-monotone fashion for
a large class of families of Hamiltonian flows.
\end{abstract}\maketitle

\section{Introduction}

The modern approach to the study of a dynamical system identifies solving the
problem with finding the global structure of phase space. On the other hand,
bifurcation theory has mostly concentrated on \emph{local behavior in the
parameter space }(so even when global phase space bifurcations are considered,
they are always considered in a small neighborhood of a given parameter
value). Here we suggest that global analysis with respect to parameters leads
to understanding of the phase space structure in parameter regimes which are
unreachable by the currently known analysis techniques. In particular, we
examine some features of the global dependence of periodically forced systems
with finite forcing amplitude on the forcing frequency.

\begin{figure}[ptb]
\begin{center}
\includegraphics[width=11cm]{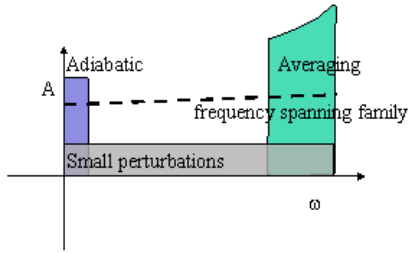}
\end{center}
\caption[Frequency spanning families]{Frequency spanning families.}%
\label{fig:spanfam}%
\end{figure}

The study of the effect of periodic forcing on homoclinic loops has been
thoroughly investigated since the times of Poincar\'{e}. Analytical
understanding of such systems has been attained in three possible limits (see
\cite{Arno88},\cite{GuHo83}): the fast and slow oscillations limits and the
small forcing limit (see figure \ref{fig:spanfam}). For fast oscillations
averaging methods apply, and together with KAM theory, these are used to
obtain upper bounds on the separatrix splitting. These bounds are
exponentially small in the frequency (\cite{Poin92},\cite{Neish84}). Lower
bounds on the separatrix splitting involve delicate analysis which has been
proved, under some structural assumptions on the flow, only in recent years
(see \cite{Ge97} and references therein). The slow oscillations regime
corresponds to the region where adiabatic theory applies. Two approaches have
been applied to this limit - the first one corresponds to extending classical
adiabatic theory to separatrix crossing (\cite{neish75},\cite{TeCaEs86}) and
the other corresponds to examining the geometrical properties of stable and
unstable manifolds to hyperbolic manifolds with slowly varying motion on them
(see \cite{KaWi90},\cite{kako94b},\cite{SoKa96},\cite{ElEs91}). Most of the
analytical studies have been performed in the small oscillations regime in
which perturbation methods apply (see (\cite{GuHo83}) and references therein).
The most important ingredient from these studies to the current paper is the
existence of a flux mechanism via lobes (see \cite{ChLe80},\cite{MaMP84}%
,\cite{Meis92},\cite{RoWi90},\cite{RkPo99}).

Here we show that for a large class of systems, the phase space structure and
its associated transport properties have some common non-monotone dependence
on the frequency of the forcing term. Namely, we establish that in some cases,
even for finite size oscillations, there exists a function which depends on
the frequency continuously. After some natural scaling, this function is
simply the sum of the areas of the incoming lobes per period, normalized by
the forcing period. Hence, results obtained in the fast and slow limits supply
information regarding the behavior in the intermediate regime where no
analytical methods apply (see figure \ref{fig:spanfam}). Furthermore, this
proposed viewpoint leads to a non-traditional scaling of the flow which is
relevant for studying its behavior near homoclinic tangles. This paper
supplies mathematical formulation and generalization of the common work of the
author with A. Poje \cite{RkPo99}, in which similar issues were considered in
the context of fluid mixing.

The paper is organized as follows: In the first section we define
the class of forced Hamiltonian families which we consider.
Roughly, these are forced Hamiltonian families with split
separatrices, for which the splitting exists for all positive
parameter values and as the family parameter varies the scaled
frequency includes the semi-infinite line. In the second section
we construct a labelling scheme for primary homoclinic orbits,
which relies on the concepts of unstable ordering introduced by
Easton \cite{East86}. This labelling scheme is valid for all
parameter values and behaves well across homoclinic bifurcations.
In the third section we prove the main result of this paper - that
the flux through the homoclinic loop depends continuously (in fact
it is $C^{1}$) in the spanning parameter of the family (in
\cite{RkPo99} we considered a restricted class of systems for
which the continuity of the flux followed immediately). To prove
this result several properties of the lobes are studied, the
details are included in appendix 2. In the fourth section we
recall the proof of RK and Poje, showing that the flux is
non-monotone in the scaled frequency and recall the implications
of these results, especially in view of their implications
regarding the stochastic zone width (see \cite{Tre98}). Section
five includes a demonstration of these concepts for the forced
center-saddle bifurcation problem. Conclusions are followed by two
appendices - in the first the homoclinic scaling is discussed
whereas in the second detailed proofs of some properties of the
lobes are included, the relation between the flux and the areas of
the exit and entry sets is explained, and the adiabatic limit is
discussed.

\section{Frequency-spanning families.}

Consider a one-and-a-half degree of freedom Hamiltonian $H(x,y,\omega
(\nu)t+\theta;\nu)$ which is $2\pi$ periodic in its last argument and depends
smoothly on the parameter $\nu\in\lbrack0,\nu^{\ast}),$ where $\nu^{\ast}>0$
may be finite or infinite. This Hamiltonian may be written in the form
\begin{equation}
H(x,y,\omega(\nu)t+\theta;\nu)=H_{0}(x,y;\nu)+H_{1}(x,y,\omega(\nu
)t+\theta;\nu) \label{ham}%
\end{equation}
where the second term has zero time average:
\[
\int_{0}^{\frac{2\pi}{\omega}}H_{1}(x,y,\omega(\nu)t+\theta;\nu)dt\equiv0
\]
and may be as large as the first one. Assume the Hamiltonian system of
(\ref{ham}):
\begin{align}
\frac{dx}{dt}  &  =\frac{\partial H(x,y,\omega(\nu)t+\theta)}{\partial
y}=u_{0}(x,y;\nu)+u_{1}(x,y,\omega(\nu)t+\theta;\nu)\label{hamsys}\\
\frac{dy}{dt}  &  =-\frac{\partial H(x,y,\omega(\nu)t+\theta)}{\partial
x}=v_{0}(x,y;\nu)+v_{1}(x,y,\omega(\nu)t+\theta;\nu)\nonumber
\end{align}
satisfies the following structural assumptions:

\begin{description}
\item [A1]$H_{0}$ and $H_{1}$ are $C^{r},r>2$ functions of all their
arguments, $H_{1}$ is $2\pi$ periodic with zero mean in its last argument .

\item[A2] For all $\nu$ values in $[0,\nu^{\ast}]$, (\ref{hamsys}) possesses a
hyperbolic periodic orbit which depends smoothly on $\nu$.

\item[A3] The family of Hamiltonians spans the frequency parameter: the
frequency $\omega(\nu)$ is a smooth $C^{r},r>2$ function, and it maps the
interval $[0,\nu^{\ast})$ onto the semi-infinite line: $[0,\infty
)=\{\omega(\nu)|\nu\in\lbrack0,\nu^{\ast})\}.$ Furthermore, $\omega
(\nu)\rightarrow0$ only at one of the interval's boundaries and at no interior point.

\item[A4] For $\omega=0$, for any phase $\theta$, at least one branch of the
stable and unstable manifolds of that periodic orbit coincide to form a
homoclinic loop. Furthermore, the manifolds and the homoclinic loops created
at $\omega=0$ depend smoothly on $\theta$, and the phase space area enclosed
by these homoclinic loops, $\ \mu(R^{L,R}(\theta))$, is non-constant and has a
finite number of extrema:
\[
\frac{d}{d\theta}\mu(R^{L,R}(\theta))\neq0\text{, for all }\theta\in
\lbrack0,2\pi]\backslash\{\theta_{1},\theta_{2},...,\theta_{N_{0}}\}\text{.}%
\]
\end{description}

\begin{figure}[ptb]
\begin{center}
\includegraphics[width=11cm]{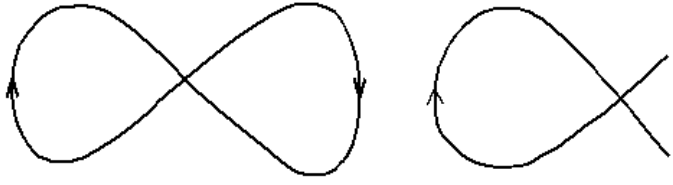}
\end{center}
\caption[Manifolds]{Manifolds geometry at $\omega=0$. \newline a. Closed
geometry (lying 8) b. Open geometry (fish).}%
\label{fig1}%
\end{figure}

For concreteness, with no loss of generality, we assume that at $\omega=0$ the
manifolds topology is of a lying figure eight (closed geometry) or of a fish
swimming to the left (open geometry), so the left branches always coincide at
$\omega=0$, see figure \ref{fig1}.

For $\omega\neq0$, consider the standard global Poincar\'{e} map
$F_{0}$ of the extended phase space $(x,y,t),$ which is simply the
time-$\frac{2\pi}{\omega}$ map of the time dependent flow (here $\theta$ is
introduced to account for phase variations):
\[
F_{\theta}:(x(0;\theta,\nu),y(0;\theta,\nu))\rightarrow(x(\frac{2\pi}{\omega
};\theta,\nu),y(\frac{2\pi}{\omega};\theta,\nu))
\]
The two dimensional symplectic map $F_{0}$ has, by the above assumptions, a
hyperbolic fixed point $\gamma(\nu)$ with its associated stable and unstable
manifolds. $p$ is a \emph{primary} homoclinic point of $\gamma(\nu)$ iff the
open segments of the unstable (respectively stable) manifold connecting
$\gamma(\nu)$ with the homoclinic point $p$, $U(\gamma(\nu),p)$ (respectively
$S(\gamma(\nu),p)$) do not intersect (\cite{East86},\cite{RoWi90}):
\[
U(\gamma(\nu),p)\cap S(\gamma(\nu),p)=\emptyset.
\]
The points $p_{i}^{k}$ in figure \ref{fig2} are all primary homoclinic points,
whereas the intersection points of the manifolds inside the shaded region in
figure \ref{fig5} are homoclinic points which are not primary. We assume:

\begin{description}
\item [A5]For all $\nu\neq0$ there is a finite number of primary homoclinic
orbits. Furthermore, for each intersecting branches of the manifolds there
exist at least two primary homoclinic orbits which are topologically transverse.

\item[A6] The number of \emph{primary} homoclinic bifurcations is bounded in
any bounded open interval of $\nu$ for all $\nu\neq0$. The order of the
\emph{primary} homoclinic bifurcations is uniformly bounded by the integer $M.$

For convenience of notation, let us assume that $M<10.$
\end{description}

Finally, to establish asymptotic behavior in the frequency, we assume there
exists an appropriate scaling so that the scaled separatrix length scales and
the maximal velocity along the separatrix are of order one:

\begin{description}
\item [A7]There exist smooth ($C^{r}$ in $\nu$) non-vanishing scaling
functions, $W_{0}(\nu),W_{1}(\nu),$ $L_{x}(\nu),L_{y}(\nu),L_{\max}(\nu),$
such that in the scaled variables:
\begin{align}
(\overline{x},\overline{y})  &  =\left(  \frac{x}{L_{x}(\nu)},\frac{y}%
{L_{y}(\nu)}\right)  ,\ \ \ \ \overline{t}=\frac{W_{0}(\nu)}{L_{\max}(\nu
)}t,\ \ \\
\ \ \ \overline{H}  &  =\frac{L_{\max}(\nu)}{W_{0}(\nu)L_{x}(\nu)L_{y}(\nu
)}H,\\
(\overline{u}_{i,}\overline{v}_{i})  &  =\frac{L_{\max}(\nu)}{W_{i}(\nu
)}\left(  \frac{u_{i}}{L_{x}(\nu)},\frac{v_{i}}{L_{y}(\nu)}\right)  ,\text{
\ }i=1,2
\end{align}
the scaled system:
\begin{align}
\frac{d\overline{x}}{d\overline{t}}  &  =\overline{u}_{0}(\overline
{x},\overline{y};\nu)+A\overline{u}_{1}(\overline{x},\overline{y}%
,\overline{\omega}\overline{t}+\theta;\nu)\label{schamsys}\\
\frac{d\overline{y}}{d\overline{t}}  &  =\overline{v}_{0}(\overline
{x},\overline{y};\nu)+A\overline{v}_{1}(\overline{x},\overline{y}%
,\overline{\omega}\overline{t}+\theta;\nu)\nonumber
\end{align}
with the two non-dimensional parameters $A,\overline{\omega}$ :
\[
A(\nu)=\frac{W_{1}(\nu)}{W_{0}(\nu)},\ \ \overline{\omega}(\nu)=\omega
(\nu)\ \frac{L_{\max}(\nu)}{W_{0}(\nu)}\ \ .
\]
satisfies assumptions A1-A6. \emph{Furthermore, for the scaled system, the
width and length of the separatrix and the maximal velocity along
it\footnote{See appendix 1 for the precise definitions, which use some of the
notation of the next section.} are all of order one for all }$\nu\in
(0,\nu^{\ast})$\emph{.}
\end{description}

$A(\nu)$ measures the relative strength of the temporal oscillations to the
mean flow whereas $\overline{\omega}(\nu)$ compares the oscillation's time
scale with the travel time along the separatrix loop (outside of the saddle
orbit neighborhood). We assert that the magnitude of these two parameters
supplies \emph{complete information on the qualitative behavior }of the system
(\ref{schamsys}) and hence on the original system. Notice that assumption A3
is now made with respect to the scaled frequency.

\begin{definition}
A family of Hamiltonian systems depending on a parameter $\nu$ which satisfies
assumptions A1-A7 is called a\emph{\ frequency-spanning homoclinic family,}
and $\nu$ is called a \emph{frequency-spanning parameter.}
\end{definition}

A few remarks are now in order:

\begin{itemize}
\item  Assumptions A1-A7 clearly hold in the standard near integrable case
(small $A(\nu)$) where the steady flow has a homoclinic loop of fixed size
(independent of $\nu$), the perturbation is generic (so A5 and A6 are
satisfied), and its frequency, $\omega(\nu)=\nu,$ spans the half real line.

\item  A5 refers only to \emph{primary }homoclinic bifurcations - otherwise it
would have been violated generically, see \cite{tu03} and references therein.

\item  In the near-integrable case the scaling functions may be simply
extracted from the integrable flow, see for example section \ref{sec:sadcen}.
In appendix 1, after some notation is established in the next section, we
propose an algorithm for defining the characteristic scales in the finite
amplitude size forcing case.

\item  The assumptions here are slightly weaker than the ones in
\cite{RkPo99}: in particular A5 and A6 replace the stronger
assumption of \cite{RkPo99} that there exists one topologically
transverse primary homoclinic orbit which depends continuously on
$\nu$, a property which may be easily violated in applications.
Relaxing this assumption requires the introduction of a new
labelling scheme for the homoclinic points which takes into
account the possible annihilation of primary homoclinic points.
\end{itemize}

Hereafter we assume the forced system is in its scaled form and we drop all
the over bars when non-ambiguous, as described next.

\section{\label{sec:label}labelling scheme for the primary homoclinic points}

Denote by $P(\nu)=$ $\{p^{r_{n}(\nu)}(\nu),n=1,...,N(\nu)\}$ the
\emph{topologically transverse primary} homoclinic \emph{orbits }of
$\gamma(\nu)$ belonging to the left branch (for simplicity of notation we
consider hereafter only the left branch of the loop), where $p^{r_{n}(\nu
)}(\nu)=\{p_{i}^{r_{n}(\nu)}(\nu)\}_{i=-\infty}^{\infty}=\{F^{i}p_{0}%
^{r_{n}(\nu)}(\nu)\}_{i=-\infty}^{\infty}$, namely the discrete index $i$
denotes iterations under the map and the index $r_{n}(\nu)$ denotes the label
of the specific orbit. Each such transverse homoclinic point, $p_{i}^{r_{n}%
}(\nu),$ defines a region, $R_{i}^{r_{n}}(\nu),$ enclosed by the closed
segments of the stable and unstable manifolds which emanate from the fixed
point and meet at $p_{i}^{r_{n}}(\nu),$ namely
\begin{equation}
\partial R_{i}^{r_{n}}(\nu)=U[\gamma(\nu),p_{i}^{r_{n}}(\nu)]\cup S[\gamma
(\nu),p_{i}^{r_{n}}(\nu)], \label{rdef}%
\end{equation}
where, clearly
\[
R_{i}^{r_{n}}(\nu)=F^{i}R_{0}^{r_{n}}(\nu).
\]

Denote by $0<\nu_{1}<\nu_{2}<....$ the ordered parameter values at
which primary homoclinic bifurcations occur (by A6 this may be
done), and by $\Upsilon$ the set of all primary homoclinic
bifurcation values in $[0,\nu^{\ast})$: $\Upsilon=\{$
$\nu_{1},\nu_{2},...\}$. In figures \ref{fig2} and \ref{fig3} \ we
illustrate a possible behavior of the manifolds before and after a
primary homoclinic bifurcation occurs. We say that
$p^{r_{n}(\nu)}$ undergoes a homoclinic bifurcation at $\nu_{i}$
if $p^{r_{n}(\nu_{1})}$ is a tangent homoclinic orbit. Next we
construct a labelling of the orbits in $P(\nu)$ which is
convenient even across homoclinic bifurcations.

\begin{lemma}
Consider a frequency spanning homoclinic family. Then, there
exists a labelling scheme for the primary homoclinic orbits
satisfying the following properties:

\begin{enumerate}
\item $r(\nu)$ is piecewise constant, changing only at values of $\nu$ at
which $p^{r(\nu)}(\nu)$ undergoes a homoclinic bifurcation.

\item  The labelling scheme respects the unstable ordering, so that $r_{n}%
(\nu)<r_{m}(\nu)$ iff $p_{0}^{r_{n}(\nu)}(\nu)<_{u}p_{0}^{r_{m}(\nu)}(\nu)$,
namely (see \cite{East86}):
\begin{equation}
U[\gamma(\nu),p_{0}^{r_{n}}(\nu)]\subset U[\gamma(\nu),p_{0}^{r_{m}}%
(\nu)]\text{ \ \ for \ \ }r_{n}<r_{m}. \label{orderhom}%
\end{equation}
\end{enumerate}
\end{lemma}

\begin{proof}
By construction. We construct such a labelling scheme inductively,
in the
intervals $(\nu_{k},\nu_{k+1}),$ starting with the interval $(\nu_{0}%
=0,\nu_{1}).$

By A5, $N(\nu)\geq2$ for all $\nu\neq0$. Since the Poincar\'{e}.
map is orientation preserving and since only topologically
transverse intersections are counted, $N(\nu)$ is even. By
assumption A6, there exists a well defined limit of the number of
primary homoclinic orbits as $\nu\rightarrow0^{+}$, so that
$N(0^{+})$ is well defined and is finite, and $\nu_{1}>0$. For
$0<\nu <\nu_{1}$, let $r_{n}(\nu)=n,n=1,...,N(0^{+}).$ The
labelling is chosen to obey the unstable ordering
$p_{0}^{i}(\nu)<_{u}p_{0}^{j}(\nu)$ for $i<j,$ where we identify:
\begin{equation}
p_{0}^{r_{N(\nu)+1}}(\nu)\equiv p_{1}^{r_{1}}(\nu)=Fp_{0}^{r_{1}}(\nu),\text{
and }p_{0}^{r_{0}}(\nu)\equiv p_{-1}^{r_{N(\nu)}}(\nu)=F^{-1}p_{0}^{r_{N(\nu
)}}(\nu). \label{cycllab}%
\end{equation}
These rules determine the labelling for $0<\nu<\nu_{1}$ up to a
cyclic permutation, and do not define the origin (which homoclinic
point is $p_{0}^{1}(\nu)$). To remove the first ambiguity, define
the labelling so that as $\nu\rightarrow0^{+}$, the first
homoclinic orbit defines the region of maximal area\footnote{By
area preservation $\mu(R_{i}^{r_{n}}(\nu))=\mu
(R_{0}^{r_{n}}(\nu))$ for all $i.$}:
$\mu(R_{0}^{1}(0^{+}))=\max_{n}(\mu (R_{0}^{n}(0^{+}))$. To remove
the second ambiguity, we define the origin ($i=0)$ as the point
for which the boundaries of \ $R_{i}^{1}(0^{+})$ are of minimal
length; Denote the arc length of the boundary of
$R_{i}^{r_{n}}(\nu)$
by $|\partial R_{i}^{r_{n}}(\nu)|$. Notice that $|\partial R_{i}^{r_{n}}%
(\nu)|\rightarrow\infty$ as $i\rightarrow\pm\infty$. Let
$p_{0}^{1}(\nu)$, satisfy $|\partial
R_{0}^{1}(\nu)|=\min_{i}|\partial R_{i}^{1}(\nu)|$. When there are
no degeneracies (here, when the maximal area is achieved at a
unique orbit and the minimal boundary length is achieved at a
unique homoclinic point) this procedure determines uniquely the
labelling of the topologically transverse homoclinic orbits for
$0<\nu<\nu_{1}.$ If a degeneracy occurs, choose any one of the
finite number of the maximizing (respectively minimizing) orbits
(respectively points). \begin{figure}[ptb]
\begin{center}
\includegraphics[width=11cm]{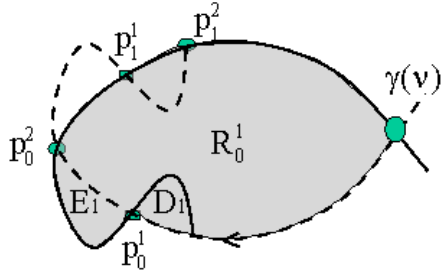}
\end{center}
\caption[Primary homoclinic point]{Primary homoclinic points. \newline Here
$N(\nu)=2,\nu<\nu_{1}$ of figure \ref{fig4}.}%
\label{fig2}%
\end{figure}

Given the labelling
$Lbl(\nu_{k-1}^{+})=\{r_{1},r_{2},...,r_{N(\nu)}\}$, for $k\geq1$,
so that $Lbl(\nu_{k-1}^{+})=Lbl(\nu)$ for $\nu\in(\nu_{k-1},\nu
_{k})$, we construct $Lbl(\nu_{k}^{+})$ as follows:

\begin{case}
If the associated homoclinic orbit, $p_{i}^{r_{n}}(\nu)$, depends continuously
on $\nu$ at $\nu_{k}$ leave $r_{n}$ unchanged.
\end{case}

\begin{case}
If $p_{i}^{r_{n}}(\nu)$ cease to exist for $\nu>\nu_{k}$ , delete $r_{n}$ from
$Lbl(\nu_{k}^{+}).$
\end{case}

\begin{case}
If $M$ additional homoclinic orbits are created between (along the unstable
ordering) $p_{i}^{r_{n}}(\nu)$ and $p_{i}^{r_{n+1}}(\nu),$ insert the $M$
labels, $r^{j}=r_{n}+j\cdot10^{-k}$, $j=1,...,M$, so that $Lbl(\nu_{k}%
^{+})=\{r_{1},r_{2},..,r_{n},r_{n}+10^{-k},r_{n}+2\cdot10^{-k},...,r_{n}%
+M\cdot10^{-k},r_{n+1},...,r_{N(\nu)}\}$
\end{case}

By construction the labelling scheme respects the unstable
ordering and does not change labels unless homoclinic bifurcation
occur, as claimed. Notice that the first case in the proof
includes two possibility which are topologically equivalent - the
first is that the homoclinic orbit $p_{i}^{r_{n}}(\nu_{k})$ is not
involved in the homoclinic bifurcation and the second possibility
is that $p_{i}^{r_{n}}(\nu_{k})$ is a tangent periodic orbit with
odd order tangency so that no topological bifurcation occurs at
$\nu_{k}$.
\end{proof}

For any $\nu$ we have constructed a finite, ordered set of labels
$Lbl(\nu)=\{r_{1},r_{2},...,r_{N(\nu)}\}$ (so $r_{1}<r_{2}<..<r_{N(\nu)}$)
with corresponding homoclinic orbits which satisfy (\ref{orderhom}). The
ordering $k(r(\nu)),r\in Lbl(\nu)$ enumerates the primary homoclinic orbits in
$P(\nu)$ by their unstable ordering. While the $r^{\prime}s$ depend smoothly
on $\nu$ as long as $p_{i}^{r}(\nu)$ has not bifurcated, $k(r(\nu))$ can be
discontinuous in $r(\nu)$ due to bifurcations of orbits $p_{i}^{r^{\prime}%
}(\nu)$ with some $r^{\prime}<r$.

For example, in the simplest, generic case of quadratic tangency, if at
$\nu_{1}$ two primary homoclinic orbits are created between (in the sense of
the unstable ordering) the $j$th and the $j+1$ homoclinic orbits which existed
for $\nu<\nu_{1}$, then these will be labeled as $r=j.1,r^{\prime}=j.2,$ so
that $p_{0}^{j}$ $<_{u}p_{0}^{j.1}<_{u}p_{0}^{j.2}<_{u}p_{0}^{j+1}(\nu),$ as
demonstrated in figure \ref{fig3}:
\begin{equation}
P(\nu)|_{\nu_{1}<\nu<\nu_{2}}=\{p^{1}(\nu),p^{2}(\nu),...,p^{j}(\nu
),p^{j.1}(\nu),p^{j.2}(\nu),p^{j+1}(\nu),...,p^{N(0^{+})}(\nu)\}.
\label{pwexm}%
\end{equation}
\begin{figure}[ptb]
\begin{center}
\includegraphics[width=11cm]{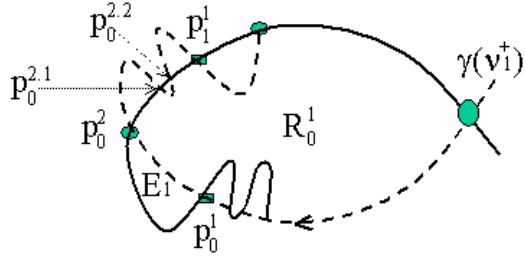}
\end{center}
\caption[labelling after the first bifurcation]{labelling after
the first bifurcation. \newline $\nu=$\underline{$\nu$} where
$\nu_{1}<$\underline{$\nu
$}$<\nu_{2}$ see figure \ref{fig4}.}%
\label{fig3}%
\end{figure}By construction, for $0<\nu<\nu_{1}$, $r_{k}=k,$ or equivalently,
$k(r;\nu)=r$. On the other hand, for $\nu_{1}<\nu<\nu_{2}$, for the generic
case of (\ref{pwexm}) we have $Lbl(\nu)|_{\nu_{1}<\nu<\nu_{2}}%
=\{1,2,..,j,j.1,j.2,j+1,...,N(0^{+})\}=\{r_{1},..,r_{N(0^{+})+2}\}$ so
\[
k(r;\nu)=\left\{
\begin{array}
[c]{l}%
r\text{ \ \ \ \ \ \ \ \ \ for }r\leq j\\
\lbrack r]+[(r-[r])\cdot10]\text{ \ \ for }j<r<j+1\\
r+2\text{ \ \ for }r\geq j+1,
\end{array}
\right.  \text{ \ \ \ \ for \ }\nu_{1}<\nu<\nu_{2}%
\]
and we see that $k(r;\nu)$ changes at $\nu_{1}$ for all $k>j,$
hence, as opposed to the $r_{n}$'s, it does not serve as a good
labelling system for the homoclinic orbits across bifurcations.
\begin{figure}[ptbptb]
\begin{center}
\includegraphics[width=11cm]{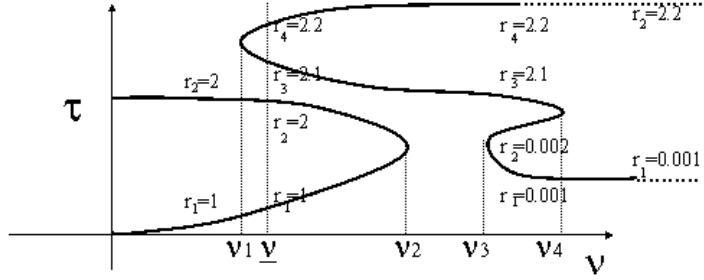}
\end{center}
\caption[Primary homoclinic bifurcation diagram (schematic)]{Primary
homoclinic bifurcation diagram (schematic). \newline $\tau$ represents natural
parametarization along the unstable manifold.}%
\label{fig4}%
\end{figure}

\section{Lobes topology}

Consider now a fixed $\nu$ value, with $Lbl(\nu)=\{r_{1},r_{2},...,r_{2n}\} $
where $N(\nu)=2n$, and $p_{0}^{r_{2n+1}}=Fp_{0}^{r_{1}}=p_{1}^{r_{1}}$.
Henceforth we omit the explicit dependence on $\nu$ unless needed.

\begin{definition}
$p^{r},p^{r^{\prime}}$ $\in P(\nu)$ are called neighbors if $|k(r)-k(r^{\prime
})|=1.$
\end{definition}

\begin{definition}
$R_{0}^{r}$ and $R_{0}^{r^{\prime}}$ (see (\ref{rdef})) are called neighboring
regions if $p^{r},p^{r^{\prime}}$ are neighbors.
\end{definition}

The difference between neighboring regions $R_{0}^{r}$ and $R_{0}^{r^{\prime}%
}$ are lobes - regions which are enclosed by the segments $U[p_{0}^{r}%
,p_{0}^{r^{\prime}}]\cup$ $S[p_{0}^{r},p_{0}^{r^{\prime}}]$:

\begin{definition}
A lobe $L(p_{i}^{r},p_{i}^{r^{\prime}})$ is the region enclosed by $\partial
L=U[p_{i}^{r},p_{i}^{r^{\prime}}]\cup$ $S[p_{i}^{r},p_{i}^{r^{\prime}}]$ where
$p_{i}^{r},p_{i}^{r^{\prime}}$ are neighboring primary homoclinic points.
\end{definition}

\subsection{Incoming and outgoing lobes}

Let
\[
D_{k}=L(p_{0}^{r_{2k-2}},p_{0}^{r_{2k-1}}),\text{ \ }E_{k}=L(p_{0}^{r_{2k-1}%
},p_{0}^{r_{2k}}),\text{ \ for }k=1,...,N(\nu)/2
\]
where (\ref{cycllab}) is used to define $D_{1}$, see figure
\ref{fig2}. Next we prove that this definition indeed corresponds
to the figure, namely that the choice of the labelling is such
that $D_{k+1}$ is exterior to $R_{0}^{r_{2k}}$ whereas $E_{k}$ is
inside the region $R_{0}^{r_{2k-1}}$:

\begin{lemma}
\label{lem:ekdk}For all $\nu>0$ and all $k$ the lobes $E_{k}$ and $D_{k}$
satisfy\footnote{$\uplus$ denotes a union of sets with disjoint interior.}:
\begin{equation}
R_{0}^{r_{2k-1}}=R_{0}^{r_{2k}}\uplus E_{k}\text{, \ \ \ \ \ \ }%
R_{0}^{r_{2k-1}}=R_{0}^{r_{2k-2}}\uplus D_{k}\text{\ \ .} \label{nghbrreg}%
\end{equation}
\end{lemma}

\begin{proof}
First we prove that for $0<\nu<\nu_{1}$, the lemma follows from
the definition of the labelling and claim \ref{uinout} which is
proved in appendix 2; recall that for these $\nu^{\prime}$s
$r_{k}=k$ and $R_{0}^{1}$ has maximal area. By the lobes and
regions definitions, it follows immediately that neighboring
regions defer from each other by lobes, so proving the lemma
amounts to proving that:

\begin{enumerate}
\item  The lobes boundaries cannot intersect the corresponding boundaries of
the regions, hence they can be either completely interior or completely
exterior to the regions.

\item  The definition of the exiting ($D$'s) and entering ($E$'s) lobes is
consistent with (\ref{nghbrreg}).
\end{enumerate}

The first part is contained in claim \ref{uinout} of appendix 2, where it is
proved that:%
\[
U(p_{0}^{r_{j}},p_{0}^{r_{j+1}})\pitchfork_{T}\partial R_{0}^{r_{j}}%
=\emptyset\text{,}%
\]%
\begin{equation}
S(p_{0}^{r_{j+1}},p_{0}^{r_{j}})\pitchfork_{T}\partial R_{0}^{r_{j+1}%
}=\emptyset, \label{eq:trans}%
\end{equation}%
\[
U(p_{0}^{r_{j}},p_{0}^{r_{j+1}})\pitchfork_{T}S(p_{0}^{r_{j+1}},p_{0}^{r_{j}%
})=\emptyset.
\]
where $\pitchfork_{T}$\ denotes topologically transverse intersection. These
results, together with the observation that $S[p_{0}^{1}(\nu),p_{0}^{2}%
(\nu)]\subset$ $\partial R_{0}^{1}(\nu)$, implies that the lobe
$E_{1}$, which is enclosed by the segments
$U[p_{0}^{1}(\nu),p_{0}^{2}(\nu)]\cup$
$S[p_{0}^{1}(\nu),p_{0}^{2}(\nu)]$ is either contained in
$R_{0}^{1}$ or is completely outside of it. Hence, either
$R_{0}^{1}=R_{0}^{2}\uplus E_{1}$ or $R_{0}^{2}=R_{0}^{1}\uplus
E_{1}$. However, the latter contradicts the choice of $R_{0}^{1}$
as the region with maximal area as $\nu\rightarrow0^{+}$. Another
manifestation of the labelling scheme may be formulated by looking
at
the manifold's orientation: $E_{1}\subset R_{0}^{1}$ iff $U[p_{0}^{1}%
(\nu),p_{0}^{2}(\nu)]$ is interior to $S[p_{0}^{1}(\nu),p_{0}^{2}(\nu)]$ as
shown in figure \ref{fig2}. The manifolds orientation is clearly preserved for
$\nu\in(0,\nu_{1})$. Furthermore, by topological transversality of the points
in $P(\nu)$ it follows immediately that all odd (respectively even) indexed
points have the same manifolds orientation as of $p_{0}^{1}(\nu)$
(respectively $p_{0}^{2}(\nu)$), so that for $0<\nu<\nu_{1}$ the lemma is
proved for all $k$.

For $\nu_{j}<\nu<\nu_{j+1}$, the lemma is proved by induction, noticing that
at the bifurcations, topologically transverse homoclinic points are
inserted/deleted in neighboring pairs, hence the orientation of the manifolds
at $p_{0}^{r_{2k+1}}(\nu)$ (respectively at $p_{0}^{r_{2k}}(\nu)$) is always
as that of $p_{0}^{1}(0^{+})$ (respectively as of $p_{0}^{2}(0^{+})$). Notice
that tangent, non topologically transverse homoclinic points are not labeled.
\end{proof}

\begin{corollary}
For all $\nu>0$ the regions with even order have smaller areas than their
neighbors, namely $R_{0}^{r_{2k+1}}$ are regions of locally maximal area
whereas $R_{0}^{r_{2k}}$ are regions of locally minimal areas.
\end{corollary}

Finally, note the following property of the lobes:

\begin{proposition}
\label{simplecon}The lobes $E_{k},D_{k}$ $k=1,...,N(\nu)$ are simply connected.
\end{proposition}

\begin{proof}
The lobe's boundaries, by definition, are given by $U(p_{0}^{r_{j}}(\nu
),p_{0}^{r_{j+1}}(\nu))\cup S(p_{0}^{r_{j}}(\nu),p_{0}^{r_{j+1}}(\nu))$, which
are both connected arcs which are joined at their end-points, hence the lobes
are connected. Each component of the boundary cannot intersect itself, and by
lemma \ref{uinout} they cannot intersect transversely each other either, so
the only source of non-trivial topology may stem from tangencies. The
appearance of primary homoclinic tangency (e.g. at $\nu=\nu_{1}$ of figure
\ref{fig3}) does not change the simple connectedness property of a lobe.
Finally, it follows from eq. (\ref{eq:trans}) that the lobe cannot reconnect
and form a ring (even in the closed geometry case!).
\end{proof}

\subsection{Continuous dependence on parameters}

For every $\nu$ there exist a collection of incoming and outgoing lobes. Their
areas changes with $\nu$, and, at homoclinic bifurcation points these changes
can be discontinuous (at even order bifurcations where lobes split or
coincide, as shown in figure \ref{fig3}). Nonetheless,

\begin{proposition}
\label{prop:c1areas}Consider a frequency-spanning homoclinic family. The sums
of the areas of the outgoing \ and incoming lobes:
\begin{equation}
\mu_{D}=\sum_{i=1}^{\frac{N(\nu)}{2}}\mu(D_{i}),\text{ \ }\mu_{E}=\sum
_{i=1}^{\frac{N(\nu)}{2}}\mu(E_{i}) \label{sumed}%
\end{equation}
are $C^{1}$ in $\nu.$
\end{proposition}

\begin{proof}
It follows from lemma \ref{simplecon} that away of the primary homoclinic
bifurcation values $\{\nu_{i}\}_{i=1}^{\infty}$ the area of each lobe depends
smoothly ($C^{r}$) on $\nu$, hence so do $\mu_{D}$ and $\mu_{E}$. By
orientation preservation, near the bifurcation values an even number of
topological transverse primary homoclinic points are added or eliminated.
Consider first a tangential bifurcation of even order occurring at $\nu_{k}$
so that $2n$ new topologically transverse homoclinic points are created at
$\nu=\nu_{k}^{+}.$ These new homoclinic points, appearing, for example,
between the homoclinic points $p^{2j},p^{2j+1}$, are denoted by $p^{2j.m}$ ,
$m=1,...,2n$ see figure 5\footnote{The other case, of homoclinic bifurcation
occuring between the homoclinic points $p^{2j-1},p^{2j}$, is similar: simply
change the corresponding indices and interchange the letters $E$ and $D$ in
this paragraph.}. The bifurcation splits the lobe $D_{j}(\nu_{1})$ (enclosed
by $U[p^{2j},p^{2j+1}]\cup S[p^{2j+1},p^{2j}]$) to the $2n+1$
disjoint\footnote{by smoothness of the manifolds} lobes $D_{j}(\nu
),E_{j+1}(\nu),D_{j+1}(\nu),...,E_{j+n}(\nu),D_{j+n}(\nu)$ created by the new
homoclinic points. As $\nu\rightarrow\nu_{k}^{+},$ by the smooth dependence of
the manifolds on $\nu$ (recall that only a finite extension of the local
stable and unstable manifolds is considered here), the areas of the interior,
newly created lobes ($E_{j+1}(\nu),...,E_{j+n}(\nu)$)\ vanish, whereas
$\mu(D_{j}(\nu))+...+\mu(D_{j+n}(\nu))\rightarrow_{\nu\rightarrow\nu_{1}^{+}%
}\mu(D_{j}(\nu_{1}))$. The area of the diminishing lobes near $\nu_{k}$ is of
the form $(\nu-\nu_{k})^{N+1/N}$ where \ $N$ is the order of the tangency at
$\nu_{k}$. Clearly the same argument applies to the case where pairs of
homoclinic orbits annihilate at $\nu_{k}.$ It follows that generically
$\mu(D)$ and $\mu(E)$ are $C^{1}$ in $\nu$ (I thank B. Fiedler for this observation).
\end{proof}

Intuitively, we think of the $D_{k}$ lobes as existing lobes and the $E_{k}$
as entering lobes. A more elegant framework for discussing transport in phase
space is achieved by introducing the notion of exit and entry sets (see
\cite{Meis92}, \cite{Meis97}):

\begin{definition}
For any region $R$, the exit ($D_{R}$) and entry ($E_{R}$) sets of $R$ are
defined as:
\[
D_{R}=\overline{R\backslash F^{-1}R},\ \ \ \ E_{R}=\overline{R\backslash FR}.
\]
\end{definition}

In appendix 2 we study the relation between entry and exit sets of the regions
$R_{0}^{r_{j}}$ and the corresponding entering and exiting lobes. In
particular, we define, for each region $R_{0}^{r_{j}}$ the set of turnstile
lobes $E^{j},D^{j\text{ }}$ in terms of union of entering and exiting lobes
(see equations (\ref{lobed}) in the appendix). We conclude that in case there
are no intersections between these two sets they correspond exactly to the
entry and exit sets of $R_{0}^{r_{j}}$. However, if such intersections exist,
the two sets are different. We provide an example which demonstrates that when
such intersections exist the measure of the entry and exit sets depends on $j.$

\emph{We therefore conclude that while the notion of entry and exit sets
appears more elegant, it has a major flaw from a dynamic point of view - it is
not invariant with respect to the regions definition.}

Another problem arising from the definition of the entry and exit sets has to
do with self intersection of the lobes (for close geometry, in the limit of
highly stretched lobes). Indeed, the boundaries of the entry and exit sets may
depend on the appearance of non-primary homoclinic orbits. The appearance of
such orbits is expected to depend sensitively on parameter values.
Equivalently, the proposition regarding the smooth dependence of the lobes
area on parameters is proved for the sum of their areas, $\mu_{E}$ (see
\ref{sumed}), and not for the area of the lobes union. In fact, any definition
of flux which depends on homoclinic orbits of higher order (namely not primary
homoclinic orbits) must be carefully examined due to the recent results on
Richness of chaos \cite{tu03} in the neighborhood of a homoclinic tangency.

Hence, we define the flux into a region using the concept of the sum of lobes areas:

\begin{definition}
Consider a frequency spanning homoclinic family. The \emph{scaled flux
function }of this family is:
\begin{equation}
\overline{f_{lux}(\nu)}=\frac{\overline{\omega(\nu)}}{2\pi}\overline{\mu_{E}%
}=\frac{\overline{\omega(\nu)}}{2\pi}\sum_{i=1}^{\frac{N(\nu)}{2}}%
\overline{\mu(E_{i})}. \label{fluxdef}%
\end{equation}
\end{definition}

\begin{corollary}
\label{cor:conflux}The flux function of a frequency spanning homoclinic family
depends continuously and has a continuous derivative (is $C^{1}$) in the
frequency spanning parameter.
\end{corollary}

\begin{proof}
By proposition \ref{prop:c1areas} and assumptions A3 and A7.
\end{proof}

Notice that the scaled system defines a scaled flux function. Substituting the
scaling function, we observe that the flux function is scaled by the typical
area and time scale associated with travelling along the homoclinic loop, as
appropriate:
\[
\overline{f_{lux}}(\nu)=\frac{\overline{\omega(\nu)}}{2\pi}\mu(\overline
{E(\nu)})=\frac{L_{\max}(\nu)}{W_{0}(\nu)}\frac{f_{lux}(\nu)}{L_{x}(\nu
)L_{y}(\nu)}=\frac{\tau_{loop}}{Area_{loop}}f_{lux}(\nu)
\]

\section{Non-monotonicity of the flux function}

\begin{theorem}
Consider a frequency-spanning homoclinic family. Then, in the scaled system
(\ref{schamsys}) the scaled flux function (eq (\ref{fluxdef})) depends
non-monotonically on the scaled frequency $\varpi(\nu).$
\end{theorem}

\begin{proof}
Recall that by corollary \ref{cor:conflux} the flux function depends
continuously on $\nu$ and that $\overline{f_{lux}(\nu)}=\frac{\overline
{\omega(\nu)}}{2\pi}\overline{\mu_{E}}.$ As $\nu\rightarrow0$, the system
(\ref{schamsys}) approaches the adiabatic limit. Then, by corollary
\ref{melimit} of Appendix 2, \ $\mu_{E}(\nu)\rightarrow\mu_{E}(0^{+})>0$,
which implies, by assumption A3 and A7, that $\overline{f_{flux}(\nu
)}\rightarrow0$. Furthermore, since $\nu_{1}$ (the first primary homoclinic
bifurcation value) is bounded away from zero, it follows from A3 that there
exists a $0<\widehat{\nu}<\nu_{1}$ such that $\overline{f_{flux}(\widehat{\nu
})}\neq0$ and $\overline{f_{flux}(\nu)}$ is monotone in the interval
$(0,\widehat{\nu}]$. On the other hand, as $\varpi(\nu)\rightarrow\infty$,
averaging of the scaled system (\ref{schamsys}) implies that the separatrix
splitting is exponentially small in $1/\varpi(\nu)$ (see \cite{Poin92}%
,\cite{Neish84},\cite{Ge97}). It follows that $\mu_{E}(\nu)\rightarrow0$
exponentially as $\nu\rightarrow\nu^{\infty}$ (where $\lim_{\nu\rightarrow
\nu^{\infty}}\varpi(\nu)=\infty$, and by A3 $\nu^{\infty}\in(0,\nu^{\ast})$),
and so does $\overline{f_{flux}(\nu)}$. In particular, there exists a
$\nu^{\prime}<$ $\nu^{\ast}$ such that for $\nu>$ $\nu^{\prime}$, $\left|
\overline{f_{flux}(\nu)}\right|  <$ $f_{flux}(\widehat{\nu})$. This proves
that $\overline{f_{flux}(\nu)}$ is non-monotone in $\nu.$
\end{proof}

We now discuss some of the implications of the non-monotone behavior of the
flux function, see \cite{RkPo99} for a fluid mixing application of these
results. First, we notice that in the near-integrable case, when the number of
primary homoclinic orbits is fixed, the flux function is proportional to the
amplitude of the Melnikov integral. Hence the above theorem proves that the
Melnikov function amplitude is non-monotone in this case, which indeed
explains the non-monotone figure one typically gets in numerous calculations
of the Melnikov integral (e.g. forced Duffing's, forced Cubic potential,
forced pendulum). It is clear that very small (large) flux corresponds to a
small (large) chaotic region near the separatrix, hence, it was suggested that
the Melnikov function amplitude gives a good characteristic to the amount of
chaos in the system. However, we proved that the flux function is non-monotone
in the spanning parameter. It is therefore natural to compare the dynamics,
and the properties of the chaotic region at equi-flux parameter values. Let
$\nu_{a}\neq\nu_{b}$ denote equi-flux parameter values such that:
\[
\overline{f_{lux}(\nu_{a})}=\overline{f_{lux}(\nu_{b})}\text{ and }\varpi
(\nu_{a})\text{ }<\varpi(\nu_{b})\text{ . }%
\]
It follows immediately, by the flux definition, that
\[
\overline{\mu_{E}(\nu_{a})}>\overline{\mu_{E}(\nu_{b})\text{ }}%
\]
namely, the sum of the lobes areas of the lower frequency is larger than that
of the higher frequency. This observation, which is well reflected when one
examines Poincar\'{e} maps of equi-flux frequencies may lead to the wrong
conclusion that adiabatic mixing leads to a larger chaotic zone. In fact, at
least for small amplitudes, we can use \cite{Tre98} to show that the converse
result hold:
\[
\overline{\mu_{stoch}}(\nu_{a})<\overline{\mu_{stoch}}(\nu_{b})
\]
where $\overline{\mu_{stoch}}$ denotes the area of the mixing zone (the area
enclosed by the KAM tori which are closest to the separatrix on either side of
the loops). A more precise statement of these observations, with numerical
demonstrations, may be found in \cite{RkPo99}. The seemingly contradicting
statements are well understood when one realizes that for small $\varpi$ the
lobes overlap considerably - in the adiabatic limit, the measure of overlap
between the turnstile lobes may approach the full lobe's area (see
\cite{RkPo99} for a proof), namely there exists a $j$ such that
\[
\mu(E^{j}\cap D^{j})\longrightarrow_{\varpi\rightarrow0}\mu(E^{j})
\]
where $E^{j},D^{j}$ are defined by eq. (\ref{lobed}) in appendix 2.

\section{\label{sec:sadcen}Forced Saddle-center bifurcation:}

As an example consider the forced saddle center bifurcation:
\begin{align*}
\overset{.}{x}  &  =p\\
\overset{.}{p}  &  =\nu-x^{2}+bx^{n}\sin\omega t.
\end{align*}
For $0<\nu\ll1$. Near $x=-\sqrt{\nu}$ a hyperbolic periodic orbit appears. Its
separatrix length scales (width and height) in the $(x,p)$ directions are
$(L_{x}(\nu),L_{p}(\nu))=(O(\sqrt{\nu}),O(\nu^{3/4}))$, so $L_{\max}=L_{x}.$
The maximal velocity along the loop scales like $W_{0}=U_{0}=O(\nu^{3/4})$.
Finally, since $W_{1}=V_{1}\frac{L_{\max}}{L_{y}}=O(\frac{b}{\sqrt[4]{\nu}})$.
we obtain the scaling:
\begin{align*}
(\overline{x},\overline{p})  &  =\left(  \frac{x}{\sqrt{\nu}},\frac{p}%
{\nu^{3/4}}\right)  ,\ \ \ \ \overline{t}=\sqrt[4]{\nu}t,\ \ \\
\ \ \ \overline{H}  &  =\nu^{-3/2}H
\end{align*}
Hence, the effective homoclinic forcing amplitude and frequency are:
\begin{equation}
A=\frac{b}{\nu^{1-n/2}},\overline{\omega}=\frac{\omega}{\sqrt[4]{\nu}}
\label{sadnodpar}%
\end{equation}
and the scaled system (with over bars dropped where non ambiguous) is:
\begin{align}
\frac{dx}{dt}  &  =p\label{eq:ressadcen}\\
\frac{dp}{dt}  &  =1-x^{2}+Ax^{n}\sin\overline{\omega}t.\nonumber
\end{align}
Namely, the effective frequency and damping increases inversely with the
bifurcation parameter. For fixed $\omega$ and fixed $A$ we recover the well
known result that the separatrix splitting is exponentially small as the
bifurcation is approached. For fixed $b$ we see that the exact dependence on
$n$ matters. For large $\overline{\omega},$ an appropriate (see \cite{Ge97})
Melnikov calculation for (\ref{eq:ressadcen}) will lead to a function of the
form:
\[
M(t;\nu,\omega,b,n)=C(\frac{\omega}{\sqrt[4]{\nu}})\frac{b}{\nu^{1-n/2}}%
\exp(-\frac{\pi\omega}{\sqrt[4]{\nu}})\cos\frac{\omega t}{\sqrt[4]{\nu}}%
\]
where $C(\cdot)$ is, up to exponential small corrections, a polynomial. If
$n=0$ and $b=\beta\nu$, then $C(\frac{\omega}{\sqrt[4]{\nu}})=K\frac{\omega
}{\sqrt[4]{\nu}}$ (see \cite{Ge97}).

Let $b=\beta\nu^{1-n/2}$, and consider a frequency $\omega(\nu)$ which
vanishes at some finite distance from the bifurcation value and is non
decreasing as the bifurcation value is approached. For example, take
$\omega=(\varepsilon-\nu)^{\alpha}\nu^{\mu}$ \ where $\beta<1$, $\varepsilon
<1,\alpha>0,-\alpha<\mu\leq0$. Then, the family of forced saddle-center
bifurcation with $\nu\in(0,\varepsilon]$ is a spanning family, satisfying
assumptions A1-A4. Verifying assumptions A5-A6 amounts to obtaining a lower
bound on the separatrix splitting for equation (\ref{eq:ressadcen}) as
$\overline{\omega}\rightarrow\infty$, a task which has been achieved in
\cite{Ge97} for the case $n=0$, and we will assume here that a similar lower
bound may be found for $n\geq0$ so that A5 and A6 are satisfied. A7 is clearly
satisfied with the scaled parameters \ref{sadnodpar}. Furthermore, the scaled
frequency $\overline{\omega}=(\varepsilon-\nu)^{\alpha}\nu^{\mu-1/4}$ is
monotone for $\nu\in(0,\varepsilon]$, and $\overline{\omega}((0,\varepsilon
])=[0,\infty)$ . In particular, taking $n=2,\alpha=1,\mu=0$ shows that the
system:
\begin{align}
\overset{.}{x}  &  =p\label{eq:fsc}\\
\overset{.}{p}  &  =\nu-x^{2}+bx^{2}\sin(\varepsilon-\nu)t\nonumber
\end{align}
is transformed to the scaled system (\ref{eq:ressadcen}) with $A=b$ and
$\overline{\omega}(\nu)=\frac{(\varepsilon-\nu)}{\sqrt[4]{\nu}}$. We therefore
conclude that the homoclinic structure of this family behaves in a
non-monotone fashion - there exists at least one value of $\nu,$ $\nu^{m}%
\in(0,\varepsilon)$ such that the scaled homoclinic flux function:
\[
\overline{f_{lux}(\nu)}=\frac{\tau_{loop}}{Area_{loop}}f_{lux}(\nu
)=\nu^{-\frac{3}{2}}f_{lux}(\nu)=\frac{1}{2\pi}\frac{(\varepsilon-\nu)}%
{\nu^{\frac{3}{2}}}\sum\mu(E_{i}(\nu))
\]
has a local extrema at $\nu^{m}$. In particular, this implies that the
corresponding Melnikov function amplitude is non-monotone in $\nu.$ One
expects that $\overline{\omega}(\nu^{m})=O(1)$, hence $\nu^{m}\approx
\varepsilon^{4}$, namely the maxima location scales with the size of the
interval on which the family is spanning.

\section{Discussion:}

We have demonstrated that the concept of spanning families, in
which one considers global dependence on parameters, leads to
non-trivial results regarding the properties of forced systems.
This approach leads to the development of a new tool - a novel way
of labelling of primary homoclinic points across bifurcations.
Using this tool we are able to prove several results regarding the
global dependence of the lobes on the spanning parameter. We
proved that lobes are simply connected. We proved that under
natural assumptions on the spanning families the flux is
continuous and non-monotone in the spanning parameter. We
demonstrated that under the same conditions the area of the exit
and entry sets may depend on the spanning parameter
discontinuously. We demonstrated that this approach leads to
non-standard view of the forced saddle-center bifurcation.

The relation of this work to higher dimensional systems, in which the spanning
parameter is in fact a slowly varying variable is yet to be explored.

\begin{acknowledgement}
Discussions with B. Fiedler and D. Turaev are greatly appreciated.
\end{acknowledgement}

\section{\label{sec:homscal}Appendix 1: Homoclinic scaling}

Below we supply an algorithm for determining the length and time scales
characteristics for the finite amplitude forcing case. Consider the original
system. Define:
\[
(l_{x}(i,n;\nu),l_{y}(i,n;\nu))=\left(  \max_{x,x^{\prime}\in\partial
R_{i}^{r_{n}}(\nu)}|x-x^{\prime}|,\max_{y,y^{\prime}\in\partial R_{i}^{r_{n}%
}(\nu)}|y-y^{\prime}|)\right)  .
\]
where the regions $R_{i}^{r_{n}}(\nu)$ are the regions defined by the primary
homoclinic points $p_{i}^{r_{n}}(\nu)$ as in section \ref{sec:label}. The
characteristic width and length of the separatrix loop are given by:%

\[
(\widehat{L}_{x}(\nu),\widehat{L}_{y}(\nu))=(l_{x}(i_{m},n_{m};\nu
),l_{y}(i_{m},n_{m};\nu)),
\]
where
\[
||(l_{x}(i_{m},n_{m};\nu),l_{y}(i_{m},n_{m};\nu))||=\min_{|i|<K,n=1,...,N(\nu
)}||(l_{x}(i,n;\nu),l_{y}(i,n;\nu))||
\]
and $K$ is some finite number. Let
\[
\widehat{L}_{\max}(\nu)=\max\{\widehat{L}_{x}(\nu),\widehat{L}_{y}(\nu)\}
\]
The choice of the labelling so that $|\partial R_{0}^{1}|$ is
minimal among all $|\partial R_{i}^{1}|$ and the oscillatory
nature of the boundary for large $i$, implies that $i_{m}$ is
expected to be small, and therefore independent of $K$ if $K$ is
sufficiently large. The characteristic velocities along the
separatrix loops are similarly defined:
\begin{align*}
\left(  \widehat{U_{i}}(\nu),\widehat{V_{i}}(\nu)\right)   &  =\left.  \left(
\lim\sup_{n,\theta,t}|u_{i}(\cdot)|,\lim\sup_{n,\theta,t}|v_{i}(\cdot
)|\right)  \right|  _{p_{0}^{r_{n}}(t;t_{0},\nu,\theta);\nu)},i=1,2\\
\widehat{W_{0}}(\nu)  &  =\widehat{L}_{\max}(\nu)\max\left\{  \frac
{\widehat{U_{0}}(\nu)}{\widehat{L}_{x}(\nu)},\frac{\widehat{V_{0}}(\nu
)}{\widehat{L}_{y}(\nu)}\right\} \\
\widehat{W}_{1}(\nu)  &  =\widehat{L}_{\max}(\nu)\max\left\{  \frac
{\widehat{U_{1}}(\nu)}{\widehat{L}_{x}(\nu)},\frac{\widehat{V_{1}}(\nu
)}{\widehat{L}_{y}(\nu)}\right\}
\end{align*}
where $p_{0}^{r_{n}}(.)$ denotes the primary homoclinic orbit with
the initial condition $p(t_{0};t_{0})=p_{0}^{r_{n}}(.)$. Since
$p_{0}^{r}$ are all primary homoclinic orbits (in particular,
bounded orbits) the functions $\widehat
{U_{i}}(\nu),\widehat{V_{i}}(\nu),\widehat{W_{i}}(\nu)$ are
clearly well defined and are independent of the choice of \ the
labelling. Away from the homoclinic bifurcation points
$\widehat{L}_{x,y}(\nu)$ depend smoothly on
$\nu$ and $\widehat{L}_{\max}(\nu),\widehat{U_{i}}(\nu),\widehat{V_{i}}%
(\nu),\widehat{W_{i}}(\nu)$ depend continuously on $\nu$. At $\nu_{j}$ these
functions may be discontinuous. Let $L_{x,y}(\nu),L(\nu),U_{i}(\nu),V_{i}%
(\nu),W_{i}(\nu)$ denote $C^{r}$- functions, which are close in the $C^{0}$
topology to the corresponding hatted functions, uniformly in $\nu$. These are
the proposed scaling functions for assumption A7.

\section{Appendix 2: Lobes}

\subsection{Neighboring regions differ by lobes}

Here we prove claim \ref{uinout} which is used to prove lemma \ref{lem:ekdk}.

\begin{claim}
\bigskip\label{uinout}For all neighboring homoclinic points $p_{0}^{r_{j}%
},p_{0}^{r_{j+1}}\in P(\nu),$ either $U(p_{0}^{r_{j}},p_{0}^{r_{j+1}})\subset
R_{0}^{r_{j}}$ or $U(p_{0}^{r_{j}},p_{0}^{r_{j+1}})\cap R_{0}^{r_{j}%
}=\emptyset$, namely
\begin{equation}
U(p_{0}^{r_{j}},p_{0}^{r_{j+1}})\pitchfork_{T}\partial R_{0}^{r_{j}}%
=\emptyset\text{ } \label{uisout}%
\end{equation}
\ and similarly$\ $%
\begin{equation}
S(p_{0}^{r_{j+1}},p_{0}^{r_{j}})\pitchfork_{T}\partial R_{0}^{r_{j+1}%
}=\emptyset\label{sisout}%
\end{equation}
(where $\pitchfork_{T}$ denotes topologically transverse intersection). In
particular,
\begin{equation}
U(p_{0}^{r_{j}},p_{0}^{r_{j+1}})\pitchfork_{T}S(p_{0}^{r_{j+1}},p_{0}^{r_{j}%
})=\emptyset. \label{uands}%
\end{equation}
\end{claim}

\begin{proof}
Recall that by definition $\partial R_{0}^{r_{j}}=U[\gamma,p_{0}^{r_{j}}]\cup
S[\gamma,p_{0}^{r_{j}}]$. Since the unstable segments cannot intersect each
other clearly $U(p_{0}^{r_{j}},p_{0}^{r_{j+1}})\cap U[\gamma,p_{0}^{r_{j}%
}]=\emptyset.$ \ So, to prove (\ref{uisout}) we need to prove that
$U(p_{0}^{r_{j}},p_{0}^{r_{j+1}})\pitchfork_{T}S[\gamma,p_{0}^{r_{j}%
}]=\emptyset$. Notice that $S[\gamma,p_{0}^{r_{j}}]=S[\gamma,p_{0}^{r_{j+1}%
}]\cup S[p_{0}^{r_{j+1}},p_{0}^{r_{j}}]$. Since $p_{0}^{r}(\nu)$ are primary
homoclinic points $U(\gamma,p_{0}^{r})$ $\cap S(\gamma,p_{0}^{r})=\emptyset$
and in particular
\begin{equation}
U(p_{0}^{r_{j}},p_{0}^{r_{j+1}})\cap S(\gamma,p_{0}^{r_{j+1}})=\emptyset
\text{,}%
\end{equation}
and
\begin{equation}
U(\gamma,p_{0}^{r_{j}})\cap S(\gamma,p_{0}^{r_{j}})=\emptyset\label{usinters}%
\end{equation}
Hence, it is left to prove (\ref{uands}). Assume the contrary: denote by $Q$
all the topologically transverse homoclinic points in this intersection.
Denote by $q\in Q$ the homoclinic point which is closest to $p_{0}^{r_{j}}$ by
the unstable ordering, namely $U(p_{0}^{r_{j}},q)\pitchfork_{T}S(p_{0}%
^{r_{j+1}},p_{0}^{r_{j}})=\emptyset$. Then, it follows from (\ref{usinters})
and the observation that $S(p_{0}^{r_{j+1}},q)\subset S(p_{0}^{r_{j+1}}%
,p_{0}^{r_{j}})\subset S(\gamma,p_{0}^{r_{j}})$ that $U(\gamma,q)\pitchfork
_{T}S(\gamma,q)=\emptyset,$ namely $q$ is a topologically transverse
\emph{primary} intersection point, which contradicts the statement that
$p_{0}^{r_{j}},p_{0}^{r_{j+1}}$ are neighbors in $P(\nu)$. Hence
(\ref{uisout}) and (\ref{uands}) are proven. To prove (\ref{sisout}) notice
that $\partial R_{0}^{r_{j+1}}=U[\gamma,p_{0}^{r_{j+1}}]\cup S[\gamma
,p_{0}^{r_{j+1}}]$, hence, to prove the claim we need to prove that
$S(p_{0}^{r_{j+1}},p_{0}^{r_{j}})\pitchfork_{T}U[\gamma,p_{0}^{r_{j+1}}]=$
$S(p_{0}^{r_{j+1}},p_{0}^{r_{j}})\pitchfork_{T}(U[\gamma,p_{0}^{r_{j}}]\cup
U(p_{0}^{r_{j}},p_{0}^{r_{j+1}}))=\emptyset.$ The intersection with the first
term is empty by (\ref{usinters}) and the transverse intersection with the
second term is empty by (\ref{uands}).
\end{proof}

\begin{corollary}
\label{correv}$U(p_{0}^{r_{j}},p_{0}^{r_{j+1}})\cap R_{0}^{r_{j}}=\emptyset$
iff $U(p_{0}^{r_{j+1}}(\nu),p_{0}^{r_{j+2}}(\nu))\subset R_{0}^{r_{j+1}}$ for
all $j=1,...,N(\nu).$
\end{corollary}

\begin{proof}
This follows from the topological transversality at $p_{0}^{r_{j+1}}(\nu)$ and
the previous lemma.
\end{proof}

\subsection{Exit and entry sets}

For any region $R$, the exit ($D_{R}$) and entry ($E_{R}$) sets of $R$ are
defined as (see \cite{Meis92}):
\[
D_{R}=\overline{R\backslash F^{-1}R},\ \ \ \ E_{R}=\overline{R\backslash FR}.
\]
where over bar denotes closure.

\begin{figure}[ptb]
\begin{center}
\includegraphics[width=5cm]{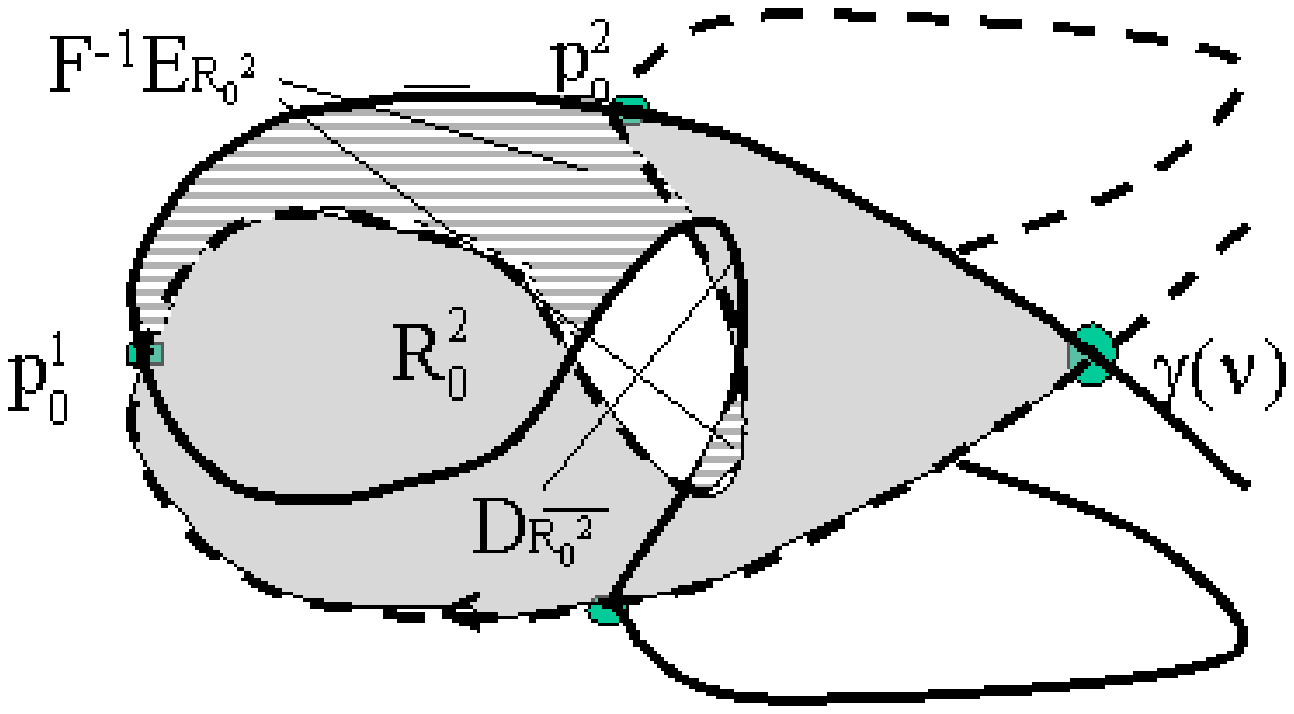} \includegraphics[width=5cm]{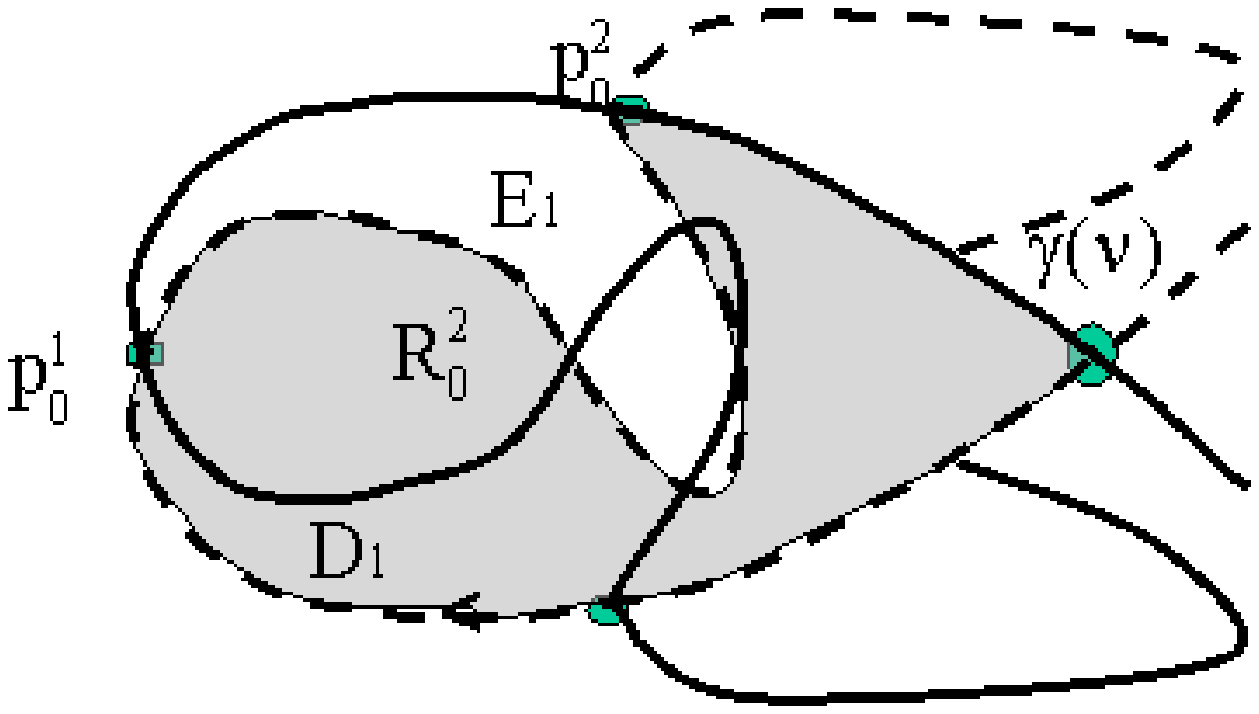}
\par
\includegraphics[width=5cm]{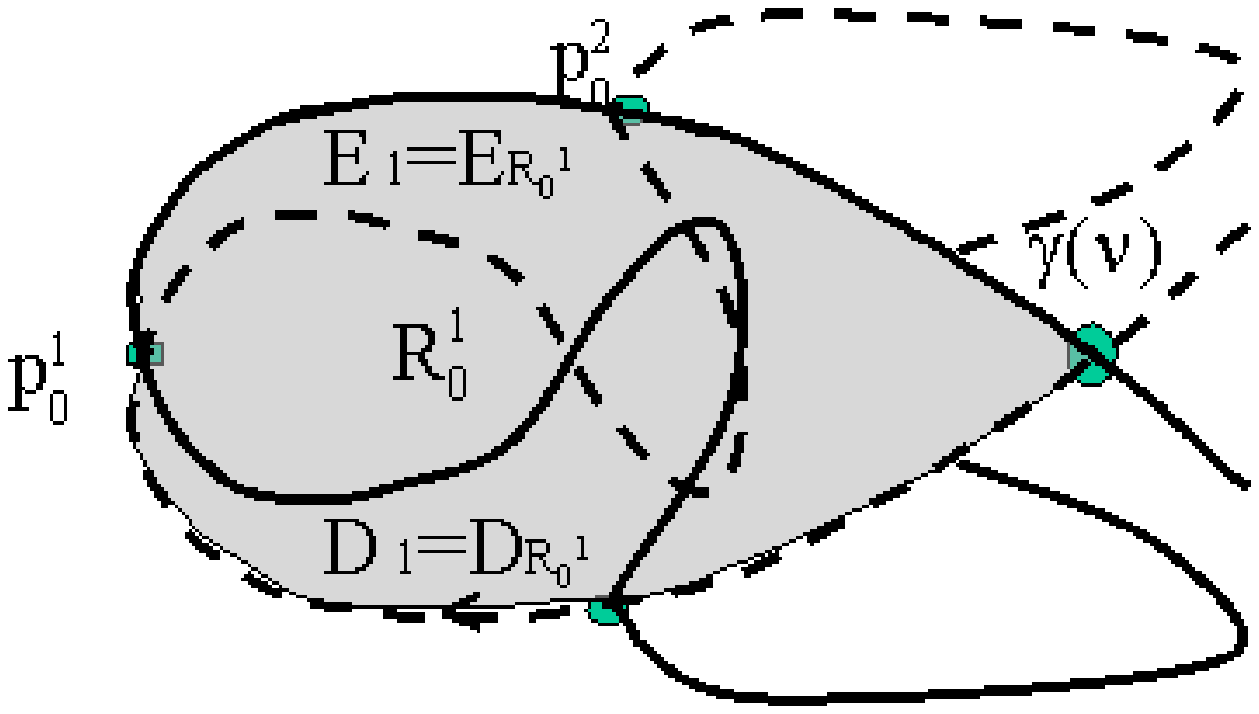}
\end{center}
\caption[Exit sets and lobe]{Exit sets and lobe.}%
\label{fig5}%
\end{figure}

Consider first the case $N(\nu)=2,$ where only two topologically transverse
primary homoclinic orbits exist. From (\ref{nghbrreg}) we immediately get
that:
\begin{align*}
R_{0}^{1}  &  =R_{0}^{2}\uplus E_{1}=F^{-1}R_{0}^{2}\uplus D_{1},\\
R_{0}^{2}  &  =R_{0}^{1}\backslash E_{1}\\
F^{-1}R_{0}^{2}  &  =R_{0}^{1}\backslash D_{1}%
\end{align*}
hence
\begin{align*}
E_{R_{0}^{1}}  &  =R_{0}^{1}\backslash FR_{0}^{1}=E_{1}\backslash FD_{1}%
=E_{1}\backslash(FD_{1}\cap E_{1})\\
D_{R_{0}^{1}}  &  =F^{-1}(FR_{0}^{1}\backslash R_{0}^{1})=F^{-1}%
(FD_{1}\backslash E_{1})=D_{1}\backslash(D_{1}\cap F^{-1}E_{1})
\end{align*}
and%
\begin{align*}
E_{R_{0}^{2}}  &  =F(F^{-1}R_{0}^{2}\backslash R_{0}^{2})=F(E_{1}\backslash
D_{1})=FE_{1}\backslash F(D_{1}\cap E_{1})\\
D_{R_{0}^{2}}  &  =R_{0}^{2}\backslash F^{-1}R_{0}^{2}=D_{1}\backslash
E_{1}=D_{1}\backslash(D_{1}\cap E_{1})
\end{align*}
therefor, we see that $D_{R_{0}^{1}}=D_{R_{0}^{2}}=D_{1}$ , $E_{R_{0}^{1}%
}=E_{1}$ and $E_{R_{0}^{2}}=FE_{1}$ iff $D_{1}\cap E_{1}=FD_{1}\cap
E_{1}=\emptyset$. In particular, in figure \ref{fig5} we demonstrate that if
$\mu(E_{1}\cap D_{1})>0$ and $E_{1}\cap FD_{1}=\emptyset$ we obtain:
\[
\mu(E_{R_{0}^{1}})=\mu(E_{1})>\mu(E_{R_{0}^{2}})=\mu(E_{1})-\mu(D_{1}\cap
E_{1}).
\]
so the measure of the entry and exit sets depend on the region one chooses,
while the measure of the entry and exit lobes are independent of this
arbitrary choice. Furthermore, we see that the entry and exit sets depend on
the nature of non-primary homoclinic orbits. The appearance of such orbits may
depend sensitively on parameter values.

For a general value of $N(\nu)$ (recall that $N(\nu)$ is even) we define the
entering and exiting lobes as:
\begin{align}
D^{2k}  &  =D^{2k-1}=\bigcup_{j=k+1}^{\frac{N(\nu)}{2}+k}D_{j}=\bigcup
_{j=k+1}^{\frac{N(\nu)}{2}}D_{j}\cup\bigcup_{j=1}^{k}FD_{j}%
,\ \ \ \label{lobed}\\
\ E^{2k}  &  =\bigcup_{j=k+1}^{\frac{N(\nu)}{2}+k}E_{j}=\bigcup_{j=k+1}%
^{\frac{N(\nu)}{2}}E_{j}\cup\bigcup_{j=1}^{k}FE_{j}\\
\ E^{2k-1}  &  =\bigcup_{j=k}^{\frac{N(\nu)}{2}+k-1}E_{j}=\bigcup_{j=k}%
^{\frac{N(\nu)}{2}}E_{j}\cup\bigcup_{j=1}^{k-1}FE_{j}%
\end{align}

It follows (after some manipulation of (\ref{nghbrreg})) that in the open
geometry case (where $E_{i}\cap E_{j}=D_{i}\cap D_{j}=\emptyset$ for all
$i\neq j$) the exit and entry sets of the regions $R_{0}^{r_{j}}$ are given
by:
\[%
\begin{array}
[c]{c}%
E_{R_{0}^{r_{j}}}=E^{j}\backslash\left(  E^{j}\cap D^{j}\right)  \\
D_{R_{0}^{r_{j}}}=F^{-1}\left(  D^{j}\backslash\left(  E^{j}\cap D^{j}\right)
\right)
\end{array}
\text{ \ if \ }E_{i}\cap E_{j}=D_{i}\cap D_{j}=\emptyset\text{ \ for all
}i\neq j.\text{\ \ \ \ \ \ \ \ \ }%
\]
In particular, in the near-integrable case, where $E^{j}\cap D^{j}=\emptyset$
for all $j=1,...,N(\nu)$ the exit and entry sets are given by the entering and
exiting lobes. Indeed, for $N(\nu)=2$ we find:%
\[
D^{2}=D^{1}=FD_{1},\text{ \ }\ E^{1}=E_{1},E^{2}=FE_{1},
\]
\bigskip and therefore the first example follows this general rule.

\subsection{Lobes in the adiabatic limit.}

The limit of $\overline{\omega}(\nu)\rightarrow0$ is called the adiabatic
limit. It corresponds to a forcing with frequency which is much smaller than
the time scale associated with the flow along the separatrix. The traditional
way of analyzing this limit is by using the adiabatic approximation. In this
approximation one considers solutions to the frozen Hamiltonian $H(x,y,\theta
;0)=H_{0}(x,y;0)+H_{1}(x,y,\theta;0)$, namely the time-dependent oscillations
are now fixed at the frozen phase $\theta$. One then studies adiabatic
variables (to leading order, the action), and asserts that in the nonlinear
case, as long as the solutions are bounded away from the separatrix, such
variables cannot change by much due to the persistence of KAM tori.
Furthermore, when the solution crosses the separatrix one can estimate the
resulting change in the adiabatic invariances and the phases by constructing
an adiabatic separatrix mapping (see for example \cite{neish75}%
,\cite{TeCaEs86},\cite{Henr82},\cite{HaYu99}). Another point of view which is
applicable to the same limit (see \cite{KaWi90},\cite{kako94b} and
\cite{SoKa96}), is concerned with the geometry of the manifolds and lobes.
Consider the extended system $(x,y,\tau=\overline{\omega}(\nu)t)$, and realize
that the manifold $\{(x,y,\tau)|x=x_{f},y=y_{f},\tau\in\lbrack0,2\pi)\}$ is a
normally hyperbolic manifold for $\overline{\omega}(\nu)=0$, hence it persists
with its stable and unstable manifolds for sufficiently small $\overline
{\omega}(\nu)$. The solutions belonging to these manifolds can be well
approximated, for semi-infinite time intervals by the extended system at
$\overline{\omega}(\nu)=0.$ Denoting the homoclinic solution to this extended
frozen system by $q_{h}(t;\tau_{0})=(x_{h}(t;\tau_{0}),y_{h}(t;\tau_{0}%
),\tau_{0})$ and the orbits belonging to the stable and unstable manifolds of
the extended system by $q^{s,u}(t;\tau_{0},\nu)=(x^{s,u}(t;\tau_{0}%
,\nu),y^{s,u}(t;\tau_{0},\nu),\tau=\overline{\omega}(\nu)t+\tau_{0})$ we see
that as $\overline{\omega}(\nu)\rightarrow0^{+}$%
\begin{align*}
q^{s}(t;\tau_{0},\nu)  &  =q_{h}(t;\tau_{0}+\overline{\omega}(\nu
)t)+...,\text{ \ \ \ }t\in(-\infty,T]\\
q^{u}(t;\tau_{0},\nu)  &  =q_{h}(t;\tau_{0}+\overline{\omega}(\nu
)t)+...,\text{ \ \ \ \ }t\in\lbrack-T,\infty)
\end{align*}
where dots stand for higher order terms in $\overline{\omega}(\nu)$, $T$ is a
positive constant and the higher order terms are small for $\overline{\omega
}(\nu)<$ $\omega^{\ast}(T)$. Recall that in the limit $\nu\rightarrow0^{+}$
(which, by A3 corresponds to the limit $\overline{\omega}(\nu)\rightarrow
0^{+}$), there are $N(0^{+})$ topologically transverse primary homoclinic
orbits, $p^{n}=p^{n}(t;\tau_{0}^{n},\nu),n=1,...,N(0^{+})$. By definition,
these orbits belong to both $q^{s}(t;\tau_{0}^{n},\nu)$ and $q^{u}(t;\tau
_{0}^{n},\nu)$ for some phase $\tau$, so that $q^{s}(0;\tau_{0}^{n},\nu)=$
$q^{u}(0;\tau_{0}^{n},\nu)$. We conclude that these phases, $\tau_{0}^{n}$,
are given, to leading order in $\overline{\omega}(\nu),$ by the phases at
which the area enclosed by the frozen homoclinic loop achieves its maxima or minima:

\begin{lemma}
(Kaper and Kovacic) Let $\{\theta_{i}\}_{i=1}^{N(0^{+})}$ denote the $\theta
$'s at the minima and maxima\footnote{i.e. disregarding any odd order
extrema.} of the homoclinic loop area $\mu(R(\theta))$, where $\mu
(R(\theta_{1}))=\max_{\theta}\mu(R(\theta))$, and $N(0^{+})\geq2$. Then, as
$\overline{\omega}(\nu)\rightarrow0^{+}$, the primary homoclinic orbits
$p^{n}(t;\tau_{0}^{n},\nu)\rightarrow q_{h}(t;\theta_{n}+\overline{\omega}%
(\nu)t)$ and:
\[
\mu(R(\theta_{n}))=\mu(R_{0}^{n})+o(1)
\]
where $R_{0}^{n}$ denotes the region enclosed by $U[\gamma,p^{n}]$ and
$S[\gamma,p^{n}]$.
\end{lemma}

\begin{proof}
See \cite{kako94b}, where it is proved that to leading order the adiabatic
Melnikov function, which measures the distance between the stable and unstable
manifolds is given, to leading order, by $\frac{d}{d\theta}\mu(R(\theta)).$
\end{proof}

\begin{corollary}
\label{melimit}As $\overline{\omega}(\nu)\rightarrow0,$%
\begin{align}
\mu(E_{k})  &  =\mu(R(\theta_{2k-1}))-\mu(R(\theta_{2k}))\nonumber\\
&  \text{\ \ \ \ \qquad
\ \ \ \ \ \ \ \ \ \ \ \ \ \ \ \ \ \ \ \ \ \ \ \ \ \ \ \ \ \ \ \ \ \ \ \ \ \ \ }%
\ \\
\mu(D_{k})  &  =\mu(R(\theta_{2k+1}))-\mu(R(\theta_{2k})),\text{
\ \ \ \ \ \ \ \ \ \ \ \ \ \ \ \ \ \ \ \ \ \ \ \ \ \ \ \ \ \ \ \ \ }\nonumber
\end{align}
hence, provided A4 is satisfied
\[
\mu_{E}(0^{+})=\lim_{\overline{\omega}(\nu)\rightarrow0^{+}}\mu_{E}(\nu
)=\sum_{k=1}^{\frac{N(0^{+})}{2}}\left(  \mu(R(\theta_{2k-1}))-\mu
(R(\theta_{2k}))\right)  >0
\]
\end{corollary}

\bigskip

The structure of the lobes in this limit is more complicated than the commonly
seen graphs of homoclinic tangles. In particular, on one hand, generically,
only one of the primary homoclinic points $p_{0}^{r_{n}}$ is bounded away from
the fixed point $\gamma(\nu)$, leading to the creation of large lobes with
area of $O(A)$ (see \cite{KaWi90}). On the other hand the lobes get stretched
and elongated, creating a web of small intersecting regions \ (see
\cite{KaWi90},\cite{ElEs91}).

Notice that if KAM tori exist, then the stable and unstable manifolds cannot
intersect them, hence they are either interior or exterior to $R_{0}^{r_{k}}$

\begin{lemma}
(Neishtadt \cite{NeChCh91}) Let $\theta_{n_{\min}}$ denote the phase at which
$\mu(R(\theta))$ is minimal ($\mu(R(\theta_{n_{\min}}))=\min\mu(R(\theta)$).
Then, as $\overline{\omega}(\nu)\rightarrow0$, the area enclosed by the
largest KAM torus inside $R_{0}^{r_{n}}$ (for any $n$), $\mu(R_{core}(\nu))$
asymptotes $\mu(R(\theta_{n_{\min}}))$:
\[
\mu(R(\theta_{n_{\min}}))=\mu(R_{core}(\nu))+...
\]
\end{lemma}

Furthermore, it is estimated in \cite{NeChCh91} that the error scales as
$O(\overline{\omega}^{2}\left|  \ln\overline{\omega}\right|  ^{2}).$

\end{document}